\documentclass[aip,reprint,author-year,nofootinbib]{revtex4-1}

\usepackage{graphicx}
\usepackage{amsmath,amssymb}             
\usepackage{stmaryrd}					
\usepackage{color}
\usepackage[dvipsnames]{xcolor}
\usepackage{color,hyperref}
\definecolor{darkblue}{rgb}{0.0,0.0,0.4}
\definecolor{red}{rgb}{0.7,0.0,0.0}
\definecolor{green}{rgb}{0.0,0.5,0.0}
\hypersetup{colorlinks,breaklinks,
            linkcolor=darkblue,urlcolor=darkblue,
            anchorcolor=darkblue,citecolor=RoyalBlue}

\def\apjl{Astrophys. J. Lett.}
\def\apj{Astrophys. J.}

\def\mnras{Mon. Not. R. Astron. Soc.}

\def\prd{Phys. Rev. D}
\def\prl{Phys. Rev. Lett.}
\def\physrep{Physics Reports}
\def\nat{Nature}

\def\aap{Astron. \& Astrophys.}

\def\jcap{Journal of Cosmology and Astroparticle Physics}
\def\na{New Astronomy}

\defcitealias{Jasche2013BORG}{{\borg} method}
\defcitealias{Jasche2015BORGSDSS}{\textsc{borg sdss}}
\defcitealias{Sutter2015VIDE}{{\vide}}
\newcommand{\vide}{\textsc{vide}}
\newcommand{\borg}{\textsc{borg}}
\begin{document}


\title{Dark matter voids in the SDSS galaxy survey}


\author{Florent Leclercq}
\email{florent.leclercq@polytechnique.org}
\affiliation{Institut d'Astrophysique de Paris (IAP), UMR 7095, CNRS -- UPMC Universit\'e Paris 6, Sorbonne Universit\'es, 98bis boulevard Arago, F-75014 Paris, France}
\affiliation{Institut Lagrange de Paris (ILP), Sorbonne Universit\'es,\\ 98bis boulevard Arago, F-75014 Paris, France}
\affiliation{\'Ecole polytechnique ParisTech,\\ Route de Saclay, F-91128 Palaiseau, France}

\author{Jens Jasche}
\affiliation{Institut d'Astrophysique de Paris (IAP), UMR 7095, CNRS -- UPMC Universit\'e Paris 6, Sorbonne Universit\'es, 98bis boulevard Arago, F-75014 Paris, France}
\affiliation{Institut Lagrange de Paris (ILP), Sorbonne Universit\'es,\\ 98bis boulevard Arago, F-75014 Paris, France}

\author{P. M. Sutter}
\affiliation{Institut d'Astrophysique de Paris (IAP), UMR 7095, CNRS -- UPMC Universit\'e Paris 6, Sorbonne Universit\'es, 98bis boulevard Arago, F-75014 Paris, France}
\affiliation{Institut Lagrange de Paris (ILP), Sorbonne Universit\'es,\\ 98bis boulevard Arago, F-75014 Paris, France}
\affiliation{Center for Cosmology and Astro-Particle Physics (CCAPP), The Ohio State University,\\ 191 West Woodruff Avenue, Columbus, OH~43210, USA}
\affiliation{National Institute for Nuclear Physics (INFN),\\ via Valerio 2, I-34127 Trieste, Italy}
\affiliation{Osservatorio Astronomico di Trieste (INAF),\\ via Tiepolo 11, I-34143 Trieste, Italy}

\author{Nico Hamaus}
\affiliation{Institut d'Astrophysique de Paris (IAP), UMR 7095, CNRS -- UPMC Universit\'e Paris 6, Sorbonne Universit\'es, 98bis boulevard Arago, F-75014 Paris, France}
\affiliation{Institut Lagrange de Paris (ILP), Sorbonne Universit\'es,\\ 98bis boulevard Arago, F-75014 Paris, France}

\author{Benjamin Wandelt}
\affiliation{Institut d'Astrophysique de Paris (IAP), UMR 7095, CNRS -- UPMC Universit\'e Paris 6, Sorbonne Universit\'es, 98bis boulevard Arago, F-75014 Paris, France}
\affiliation{Institut Lagrange de Paris (ILP), Sorbonne Universit\'es,\\ 98bis boulevard Arago, F-75014 Paris, France}
\affiliation{Department of Physics, University of Illinois at Urbana-Champaign,\\ 1110 West Green Street, Urbana, IL~61801, USA}
\affiliation{Department of Astronomy, University of Illinois at Urbana-Champaign,\\ 1002 West Green Street, Urbana, IL~61801, USA}


\date{\today}

\begin{abstract}
\noindent What do we know about voids in the dark matter distribution given the Sloan Digital Sky Survey (SDSS) and assuming the $\Lambda$CDM model? Recent application of the Bayesian inference algorithm {\borg} to the SDSS Data Release 7 main galaxy sample has generated detailed Eulerian and Lagrangian representations of the large-scale structure as well as the possibility to accurately quantify corresponding uncertainties. Building upon these results, we present constrained catalogs of voids in the Sloan volume, aiming at a physical representation of dark matter underdensities and at the alleviation of the problems due to sparsity and biasing on galaxy void catalogs. To do so, we generate data-constrained reconstructions of the presently observed large-scale structure using a fully non-linear gravitational model. We then find and analyze void candidates using the \textsc{vide} toolkit. Our methodology therefore predicts the properties of voids based on fusing prior information from simulations and data constraints. For usual void statistics (number function, ellipticity distribution and radial density profile), all the results obtained are in agreement with dark matter simulations. Our dark matter void candidates probe a deeper void hierarchy than voids directly based on the observed galaxies alone. The use of our catalogs therefore opens the way to high-precision void cosmology at the level of the dark matter field. We will make the void catalogs used in this work available at \href{http://www.cosmicvoids.net}{http://www.cosmicvoids.net}.
\end{abstract}


\maketitle



\section{Introduction}

Observations of the cosmic large-scale structure (LSS) have revealed that galaxies tend to lie in thin wall-like structures surrounding large underdense regions known as voids, which constitute most of the volume of the Universe. Although the discovery of cosmic voids dates back to some of the first galaxy redshift surveys \citep{Gregory1978,Kirshner1981,deLapparent1986} and their significance was assessed in some early studies \citep{Martel1990,vandeWeygaert1993,Goldberg2004}, the systematic analysis of void properties has only been considered seriously as a source of cosmological information in the last decade \citep[e.g.][and references therein]{Sheth2004,Colberg2005,Viel2008,BetancortRijo2009,Lavaux2010,Biswas2010,vandeWeygaert2011,Lavaux2012}. Like overdense tracers of the density field such as clusters, voids can be studied by statistical methods in order to learn about their distribution and properties compared to theoretical predictions.

Generally, direct sensitivity of void statistics to cosmology is only guaranteed for the underdense regions of the overall matter density field, which includes a large fraction of dark matter (DM). These are the physical voids in the LSS, for which theoretical modeling is established. However, absent direct measurements of dark matter underdensities, current void catalogs are defined using the locations of galaxies in large redshift surveys \citep{Pan2012,Sutter2012DR7CATALOGS,Sutter2014DR9CATALOGS,Nadathur2014a}. Since galaxies trace the underlying mass distribution only sparsely, void catalogs are subject to uncertainty and noise. Additionally, numerical simulations show that there exists a population of particles in cosmic voids. This is an indication of physical biasing in galaxy formation: there is primordial dark and baryonic matter in voids, but due to the low density, little galaxy formation takes place there. Additionally, due to complex baryonic physics effects during their formation and evolution, galaxies are biased tracers of the underlying density field, which gives rise to qualitatively different void properties.

The sensitivity of void properties to the sampling density and biasing of the tracers has only been recently analyzed in depth on simulations, by using synthetic models to mimic realistic surveys. \cite{Little1994,Benson2003,Tinker2009,Sutter2014SPARSITYBIAS} found that the statistical properties of voids in galaxy surveys are not the same as those in dark matter distributions. At lower tracer density, small voids disappear and the remaining voids are larger and more spherical. Their density profiles get slightly steeper, with a considerable increase of their compensation scale, which potentially may serve as a static ruler to probe the expansion history of the Universe \citep{Hamaus2014a}. \cite{Hamaus2014b} recently proposed a universal formula for the density profiles of voids, describing in particular dark matter voids in simulations \citep[see also][]{Colberg2005,Paz2013,Ricciardelli2014,Nadathur2014b}. The connection between galaxy voids and dark matter voids on a one-by-one basis is difficult due to the complex internal hierarchical structure of voids \citep{Dubinski1993,vandeWeygaert1993,Sahni1994,Sheth2004,Aragon-Calvo2013,Sutter2014DR9CATALOGS,Sutter2014DMGALAXYVOIDS}. However, the nature of this relationship determines the link between a survey, with its particular tracer density, and the portion of the cosmic web that it represents. Understanding this connection is of particular importance in light of recent results which probe the LSS via its effect on photons geodesics. These results include \cite{Melchior2014,Clampitt2014}, which probe the dark matter distribution via weak gravitational lensing; \cite{Ilic2013,PlanckCollaboration2014ISW} for the detection of the integrated Sachs-Wolfe effect in the cosmic microwave background, sensitive to the properties of dark energy. As a response to this demand, \cite{Sutter2014DMGALAXYVOIDS} found that voids in galaxy surveys always correspond to underdensities in the dark matter, but that their centers may be offset and their size can differ, in particular in sparsely sampled surveys where void edges suffer fragmentation.

While previous authors offer broad prescriptions to assess the effects of sparsity and biasing of the tracers on voids, the connection between galaxy voids of a particular survey and dark matter underdensities remains complex. In particular, disentangling these effects from cosmological signals in presence of the uncertainty inherent to any cosmological observation (selection effects, survey mask, noise, cosmic variance) remains an open question. In this work, we propose a method designed to circumvent the issues due to the conjugate and intricate effects of sparsity and biasing on galaxy void catalogs. In doing so, we will show that voids in the dark matter distribution can be constrained by the \textit{ab initio} analysis of surveys of tracers, such as galaxies. We will demonstrate the feasibility of our method and obtain catalogs of dark matter voids candidates in the Sloan Digital Sky Survey Data Release 7.

Our method is based on the identification of voids in the dark matter distribution inferred from large-scale structure surveys. The constitution of such maps from galaxy positions, also known as ``reconstruction'', is a field in which Bayesian methods have led to enormous progress over the last few years. Initial approaches typically relied on approximations such as a multivariate Gaussian or log-normal distribution for density fields, with a prescription for the power spectrum to account for the correct two-point statistics \citep{Lahav1994,Zaroubi2002,Erdovgdu2004,Kitaura2008,Kitaura2009,Kitaura2010,JascheKitaura2010,Jasche2010a,Jasche2010b}. However, due to their potentially complex shapes, proper identification of structures such as voids requires reconstructions correct not only at the level of the power spectrum, but also higher-order correlators. Inferences of this kind from observational data have only been made possible very recently by the introduction of physical models of structure formation in the likelihood. This naturally moves the problem to the inference of the initial conditions from which the large-scale structure originates \citep{Jasche2013BORG,Kitaura2013,Wang2013}.

This work exploits the recent application of the {\borg} (Bayesian Origin Reconstruction from Galaxies, \citealp[\citetalias{Jasche2013BORG} hereafter]{Jasche2013BORG}) algorithm to the Sloan Digital Sky Survey (SDSS) galaxies \citep[\citetalias{Jasche2015BORGSDSS} hereafter]{Jasche2015BORGSDSS}, and on the subsequent generation of constrained non-linear realizations of the present large-scale distribution of dark matter. {\borg} is a full-scale Bayesian framework, permitting the four-dimensional physical inference of density fields in the linear and mildly non-linear regime, evolving gravitationally from the initial conditions to the presently observed large-scale structure. By exploring a highly non-linear and non-Gaussian LSS posterior distribution via efficient Markov Chain Monte Carlo methods, it also provides naturally and fully self-consistently accurate uncertainty quantification for all derived quantities. A straightforward use of reconstructed initial conditions is to resimulate the considered volume \citep{Lavaux2010a,Kitaura2013,Hess2013}. In the same spirit, building upon the inference of the initial conditions by {\borg}, one can generate a set of data-constrained realizations of the present large-scale structure via full $N$-body dynamics. As we will show, we make use of initial conditions reconstructed by {\borg} without any further post-processing, which demonstrates the high quality of inference results.

Due to the limited number of phase-space foldings, the influence of non-linearity in cosmic voids is expected to be milder as compared to galaxies and dark matter halos (\citealp{Neyrinck2012,Neyrinck2013b,Leclercq2013}, see also \citealp{Abel2012,Falck2012,Shandarin2012}). For this reason, voids are more closely related to the initial conditions of the Universe, which makes them the ideal laboratories for physical application of Bayesian inference with {\borg}. In this work, we apply the void finder algorithm {\vide} \citep[\citetalias{Sutter2015VIDE} hereafter]{Sutter2015VIDE}, based on \textsc{zobov} \citep{Neyrinck2008}, to data-constrained, non-linear reconstructions of the LSS. Each of them is a full physical realization of densely-sampled particles tracing the dark matter density field. In this fashion, we construct catalogs of dark matter voids in the SDSS volume robust to sparsity and biasing of galaxies. As we will show, this procedure drastically reduces statistical uncertainty in void catalogs. Additionally, the use of data-constrained reconstructions allows us to extrapolate the void identification in existing data (e.g. at very small or at the largest scales, at high redshift or near the survey boundary).

As described in \citetalias{Jasche2013BORG,Jasche2015BORGSDSS}, the {\borg} inference framework possesses a high degree of control on observational systematic and statistical uncertainties such as noise, survey geometry and selection effects. Uncertainty quantification is provided via efficient sampling of the corresponding LSS posterior distribution. The resultant set of initial and final density field realizations yields a numerical representation of the full posterior distribution, capturing all data constraints and observational uncertainties. Building upon these results, in this work, we will extend our Bayesian reasoning to void catalogs. Specifically, we apply full non-linear $N$-body dynamics to a set of data-constrained initial conditions to  arrive at a set of non-linear dark matter density fields at the present epoch. As a result, we obtain a probabilistic description of non-linear density fields constrained by SDSS observations. Applying the {\vide} void finder to this set of reconstructions yields $N$ data-constrained realizations of the catalog, representing the posterior probability distribution for dark matter voids given observations. In this fashion, we have fully Bayesian access to uncertainty quantification via the variation between different realizations. In particular, we are now able to devise improved estimators for any void statistics by the use of Blackwell-Rao estimators. To assess the robustness of this technique for cosmological application, we focus on three key void observables: number functions, ellipticity distributions and radial density profiles. These are especially sensitive probes of non-standard cosmologies \citep{Bos2012} and are well understood in both data and simulations \citep[e.g][]{Sutter2014DR9CATALOGS}.

As a general matter, we stress that these data-constrained realizations of dark matter void catalogs were obtained assuming a $\Lambda$CDM prior. Using our products for model testing therefore requires care: in the absence of data constraints, one will simply be dealing with realizations of the $\Lambda$CDM prior. Consequently, any departure from unconstrained $\Lambda$CDM predictions are driven by the data. Conversely, for model tests where the data are not strongly informative, agreement with $\Lambda$CDM is the default answer.

This paper is organized as follows. In section \ref{sec:Methodology}, we describe our methodology: Bayesian inference with the {\borg} algorithm, non-linear filtering of the results, void identification technique and Blackwell-Rao estimators for void statistics. In section \ref{sec:Properties of dark matter voids}, we examine the properties of the dark matter voids in our catalogs. Finally, in section \ref{sec:Summary and conclusions} we summarize our results, discuss perspectives for existing and upcoming galaxy surveys and offer concluding comments. 


\section{Methodology}
\label{sec:Methodology}

In this section, we describe our methodology step by step:

\begin{enumerate}
\item {inference of the initial conditions with {\borg} (section \ref{sec:Bayesian large-scale structure inference with the BORG algorithm}),}
\item {generation of data-constrained realizations of the SDSS volume (section \ref{sec:Generation of data-constrained reconstructions}),}
\item {void finding and processing (section \ref{sec:Void finding and processing}),}
\item {combination of different void catalogs with Blackwell-Rao estimators (section \ref{sec:Blackwell-Rao estimators for dark matter void realizations}).}
\end{enumerate}

In section \ref{sec:Galaxy void catalog and dark matter simulation}, we describe the void catalogs used as references for comparison with our results. These are galaxy void catalogs directly based on SDSS galaxies without use of our methodology, and catalogs of voids in dark matter simulations.

\begin{figure} 
  \centering
  {Standard galaxy void identification\\
  \includegraphics[type=pdf,ext=.pdf,read=.pdf,width=\columnwidth]{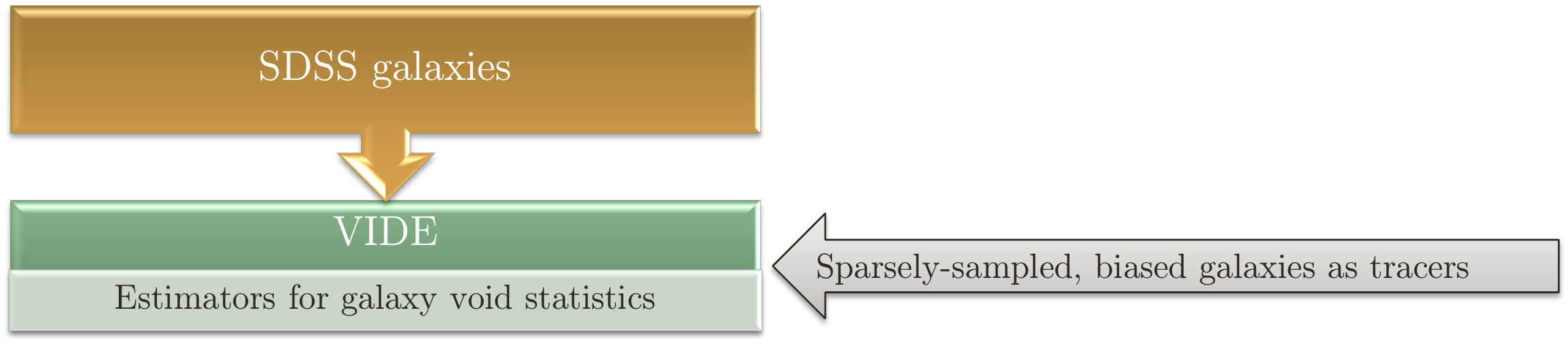}\vspace{15pt}\\
  Inference of dark matter voids\\\vspace{2pt}
  \includegraphics[type=pdf,ext=.pdf,read=.pdf,width=\columnwidth]{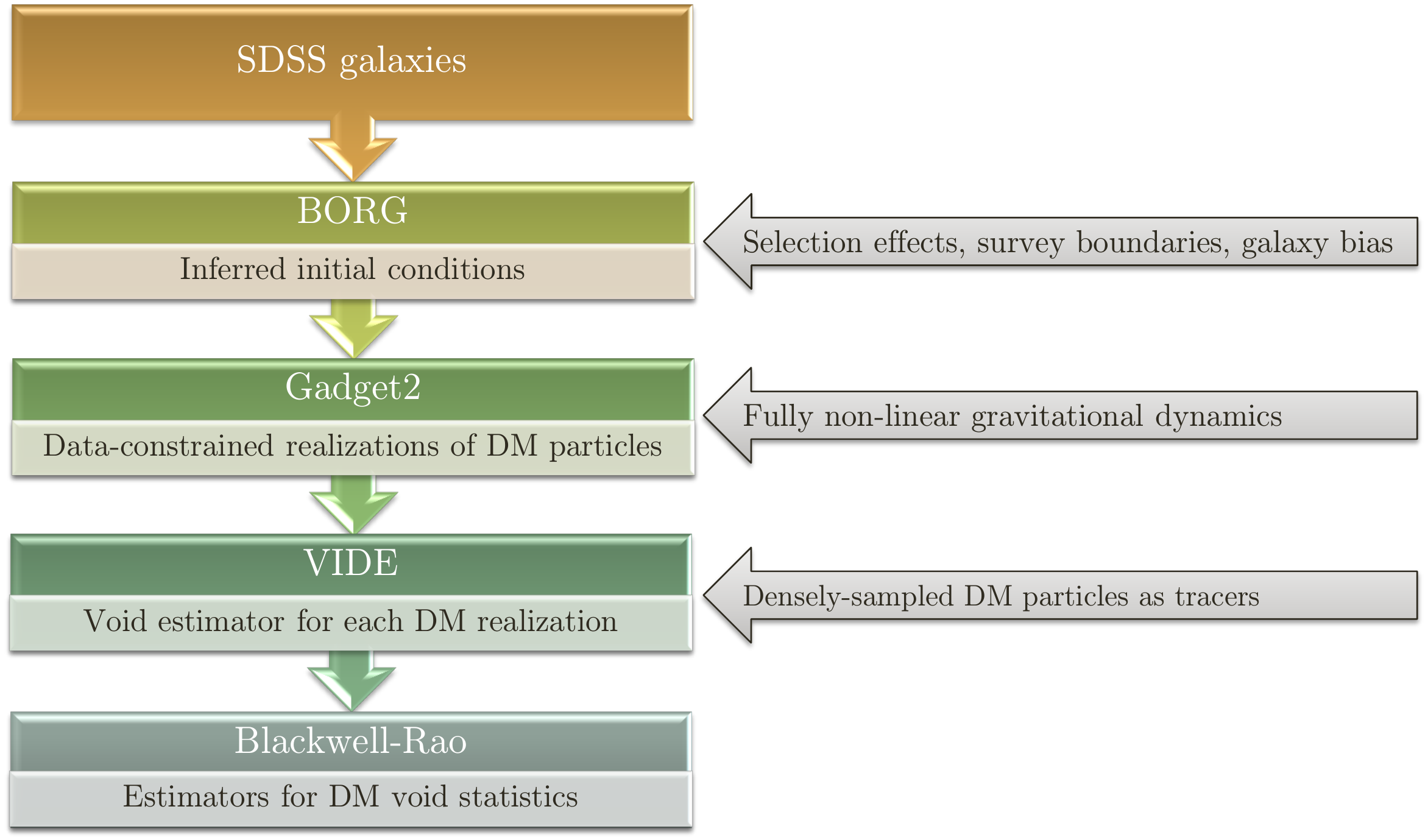}}
  \caption{
  Schematic representation of our methodology for the inference of dark matter voids (lower panel) in comparison to the standard approach for the identification of galaxy voids (upper panel).
           }
\label{fig:method}
\end{figure}

A schematic representation of our procedure is represented in figure \ref{fig:method}, in comparison to the standard approach of finding voids using galaxies as tracers.

\subsection{Bayesian large-scale structure inference with the BORG algorithm}
\label{sec:Bayesian large-scale structure inference with the BORG algorithm}

This works builds upon previous results, obtained by the application of {\borg} \citepalias[Bayesian Origin Reconstruction from Galaxies,][]{Jasche2013BORG} to SDSS main galaxy data \citepalias{Jasche2015BORGSDSS}. The {\borg} algorithm is a fully probabilistic inference machinery aiming at the analysis of linear and mildly-non-linear density and velocity fields in galaxy observations. It incorporates a physical model of cosmological structure formation, which translates the traditional task of reconstructing the non-linear three-dimensional density field into the task of inferring corresponding initial conditions from present cosmological observations. This approach yields a highly non-trivial Bayesian inference, requiring to explore very high-dimensional and non-linear spaces of possible solutions to the initial conditions problem from incomplete observations. Typically, these parameter spaces comprise on the order of $10^6$ to $10^7$ parameters, corresponding to the elements of the discretized observational domain.

Specifically, the {\borg} algorithm explores a posterior distribution consisting of a Gaussian prior, describing the statistical behavior of the initial density field at a cosmic scale factor of $a=10^{-3}$,  linked via second-order Lagrangian perturbation theory (2LPT) to a Poissonian model of galaxy formation at the present epoch (for details see \citetalias{Jasche2013BORG} and \citetalias{Jasche2015BORGSDSS}). As pointed out by previous authors \cite[see e.g.][]{Moutarde1991,Buchert1994,Bouchet1995,Scoccimarro2000,Bernardeau2002,Scoccimarro2002}, 2LPT describes the one, two and three-point statistics correctly and represents higher-order statistics very well. Consequently, the {\borg} algorithm naturally accounts for features of the cosmic web, such as filaments, that are typically associated to high-order statistics induced by non-linear gravitational structure formation processes.

Besides physical structure formation, the posterior distribution also accounts for survey geometry, selection effects and noise, inherent to any cosmological observation. Corresponding full Bayesian uncertainty quantification is provided by exploring this highly non-Gaussian and non-linear posterior distribution via an efficient Hamiltonian Markov Chain Monte Carlo sampling algorithm (see \citetalias{Jasche2013BORG} for details). In order to account for luminosity dependent galaxy bias \citep{Jasche2013BIAS} and to make use of automatic noise calibration, we further use modifications introduced to the original {\borg} algorithm by \citetalias{Jasche2015BORGSDSS}.

In this work, we make use of the 12,000 samples of the posterior distribution generated by \citetalias{Jasche2015BORGSDSS}, which constitute highly-detailed and accurate reconstructions of the initial and present-day density fields constrained by SDSS observations.

\subsection{Generation of data-constrained reconstructions}
\label{sec:Generation of data-constrained reconstructions}

\begin{figure*} 
  \centering
  {\includegraphics[type=pdf,ext=.pdf,read=.pdf,width=\textwidth]{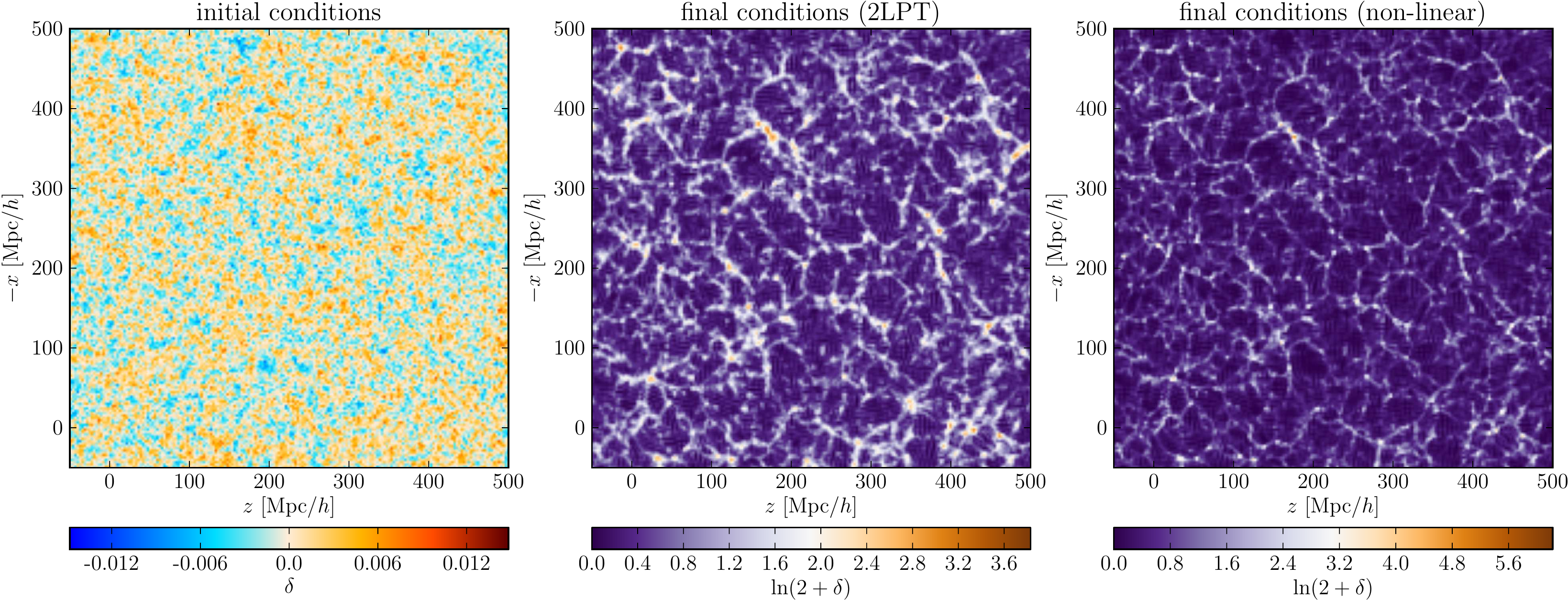}}
  \caption{
  Non-linear filtering of {\borg} results. Slices through one sample of initial (left panel) and final density fields (middle panel) inferred by {\borg}. The final density field (middle panel) is a prediction of the 2LPT model used by {\borg}. On the right panel, a slice through the data-constrained realization obtained with the same sample via non-linear filtering (fully non-linear gravitational structure formation starting from the same initial conditions) is shown. 
           }
\label{fig:filtering}
\end{figure*}

We rely on a subset of statistically independent initial conditions realizations, provided by \citetalias{Jasche2015BORGSDSS}, to generate a set of data-constrained realizations of the present large-scale structure. The initial density field, defined on a cubic equidistant grid with side length of 750~Mpc/$h$ and $256^3$ voxels, is populated by $512^3$ dark matter particles placed on a regular Lagrangian grid. The particles are evolved with 2LPT to the redshift of $z=69$, followed by a propagation with the non-linear \textsc{gadget-2} cosmological code \citep{Springel2001,Springel2005} from $z=69$ to $z=0$. In this fashion, we generate fully non-linear, data-constrained reconstructions of the present-day large-scale dark matter distribution. For this work, we use an ensemble of 11 such reconstructions.

The final conditions inferred by {\borg} are accurate only at linear and mildly non-linear scales. Application of fully non-linear dynamics to the corresponding initial conditions acts as an additional filtering step, extrapolating predictions to unconstrained non-linear regimes. In a Bayesian approach, this new information can then be tested with complementary observations in the actual sky for updating our knowledge on the Universe.

An illustration\footnote{In figures \ref{fig:filtering} and \ref{fig:slice_voids}, we kept the coordinate system of \citetalias{Jasche2015BORGSDSS}.} of the non-linear filtering procedure is presented in figure \ref{fig:filtering}. By comparing initial and final density fields, one can see correspondences between structures in the present Universe and their origins. Comparing the final density fields before and after filtering (middle and left panels), one can check the conformity of the linear and mildly non-linear structures at large and intermediate scales, correctly predicted by 2LPT. Small-scale structures, corresponding to the deeply non-linear regime, are much better represented after non-linear filtering (resulting particularly in sharper filaments and clusters). $N$-body dynamics also resolves much more finely the substructure of voids -- known to suffer from spurious artifacts in 2LPT, namely the presence of peaky, overdense spots where there should be deep voids \citep{Neyrinck2013a,Leclercq2013} -- which is of relevance for the purpose of this work.

\subsection{Void finding and processing}
\label{sec:Void finding and processing}

\subsubsection{Void finding}
\label{sec:Void finding}

We identify and post-process voids with the {\vide} (Void IDentification and Examination) toolkit\footnote{\href{http://www.cosmicvoids.net}{http://www.cosmicvoids.net}}~\citepalias{Sutter2015VIDE}, which uses a highly modified version of \textsc{zobov} \citep{Neyrinck2008,Lavaux2012,Sutter2012DR7CATALOGS} to create a Voronoi tessellation of the tracer particle population and the watershed transform to group Voronoi cells into zones and voids \citep{Platen2007}. The watershed transform identifies catchment basins as the cores of voids, and ridgelines, which separate the flow of water, as the boundaries of voids. It naturally builds a nested hierarchy of voids \citep{Lavaux2012,Bos2012}. For the purposes of this work, we examine all voids regardless of their position in the hierarchy. The pipeline imposes a density-based threshold within the void finding operation: voids only include as additional members Voronoi zones if the minimum ridge density between that zone and the void is less than 0.2 times the mean particle density (\citealt{Platen2007}; see \citealt{Blumenthal1992,Sheth2004} for the role of the corresponding $\delta = -0.8$ underdensity). If a void consists of only a single zone (as they often do in sparse populations) then this restriction does not apply. 

{\vide} provides several useful definitions used in this work,  such as the effective radius,
\begin{equation}
R_\mathrm{v} \equiv \left( \frac{3}{4\pi} V \right)^{1/3} ,
\end{equation}
where $V$ is the total volume of the Voronoi cells that contribute to the void. We use this radius definition to ignore voids with $R_\mathrm{v}$ below the mean particle spacing $\bar{n}^{-1/3}$ of the tracer population, as these are increasingly affected by Poisson fluctuations. {\vide} also reports the volume-weighted center, or macrocenter, as
\begin{equation}
\textbf{x}_\mathrm{v} \equiv \frac{1}{\sum_i V_i} \sum_i \textbf{x}_i V_i,
\end{equation}
where $\textbf{x}_i$ and $V_i$ are the positions and Voronoi volumes of each tracer particle $i$, respectively.

In each tracer population, the {\vide} pipeline provides void estimators ; in particular, the three statistics we will focus on in section \ref{sec:Properties of dark matter voids}: number count, ellipticity distribution and radial density profile. 

\begin{figure*} 
  \centering
  {\includegraphics[type=pdf,ext=.pdf,read=.pdf,width=\columnwidth]{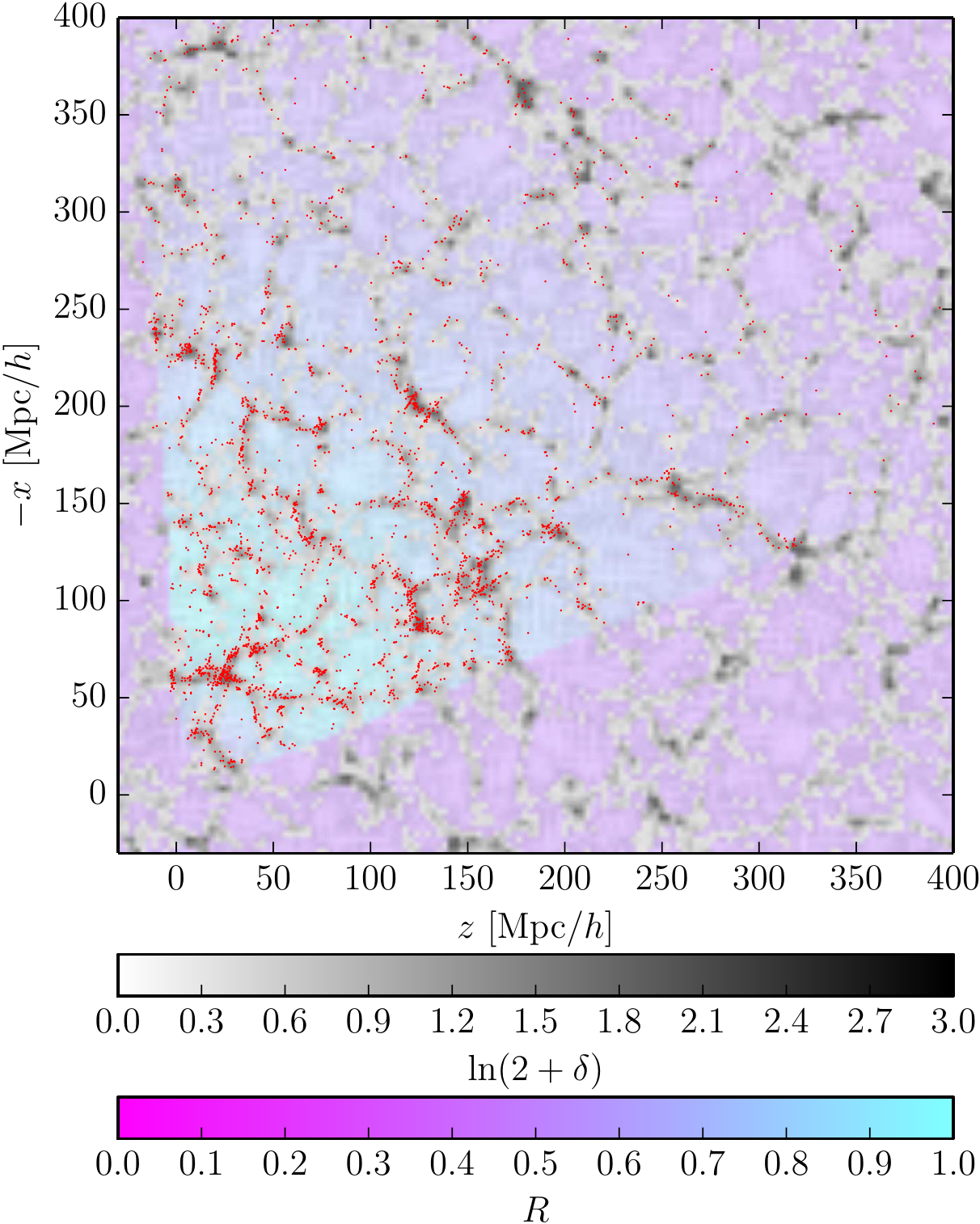}}
  {\includegraphics[type=pdf,ext=.pdf,read=.pdf,width=\columnwidth]{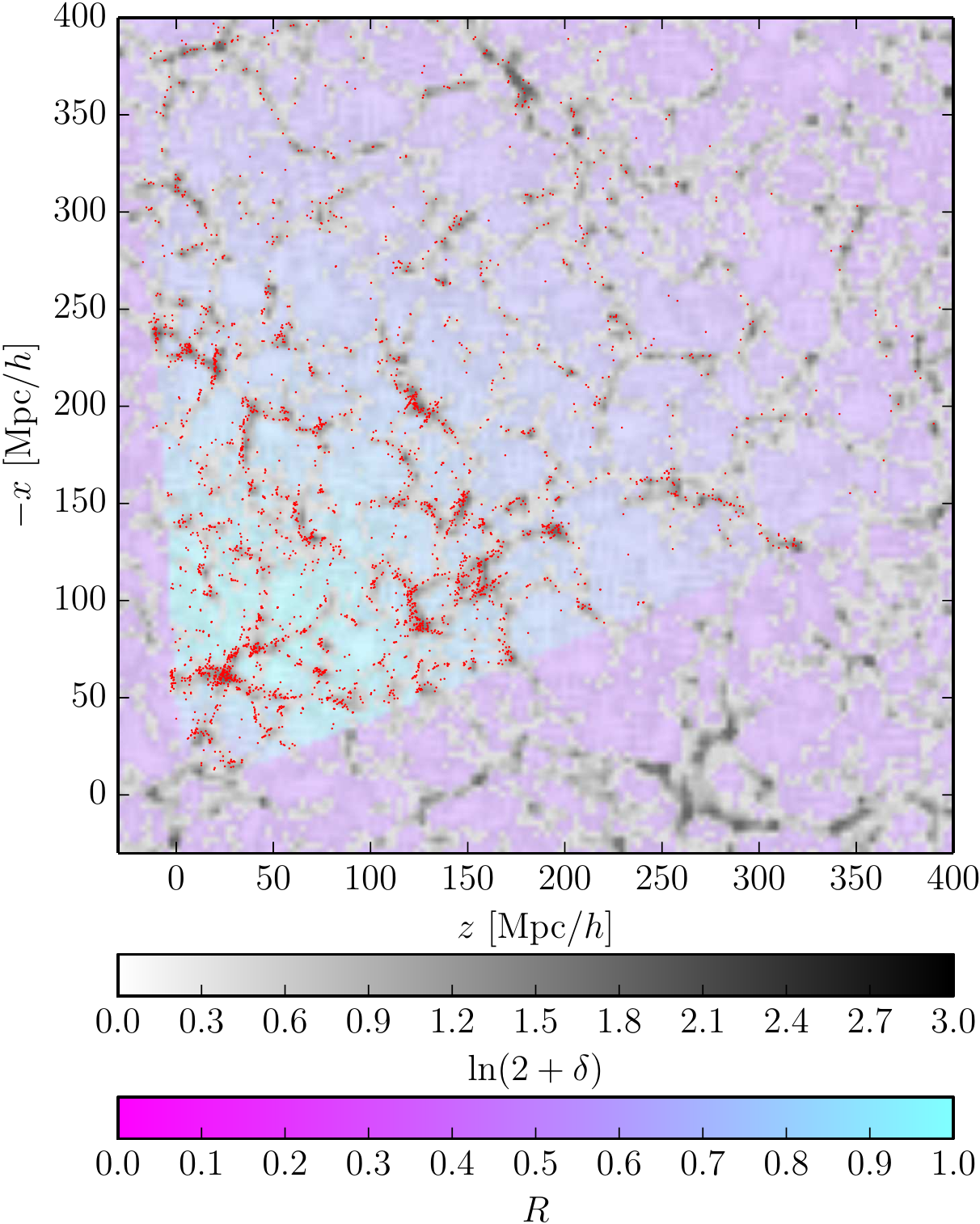}}
  {\includegraphics[type=pdf,ext=.pdf,read=.pdf,width=\columnwidth]{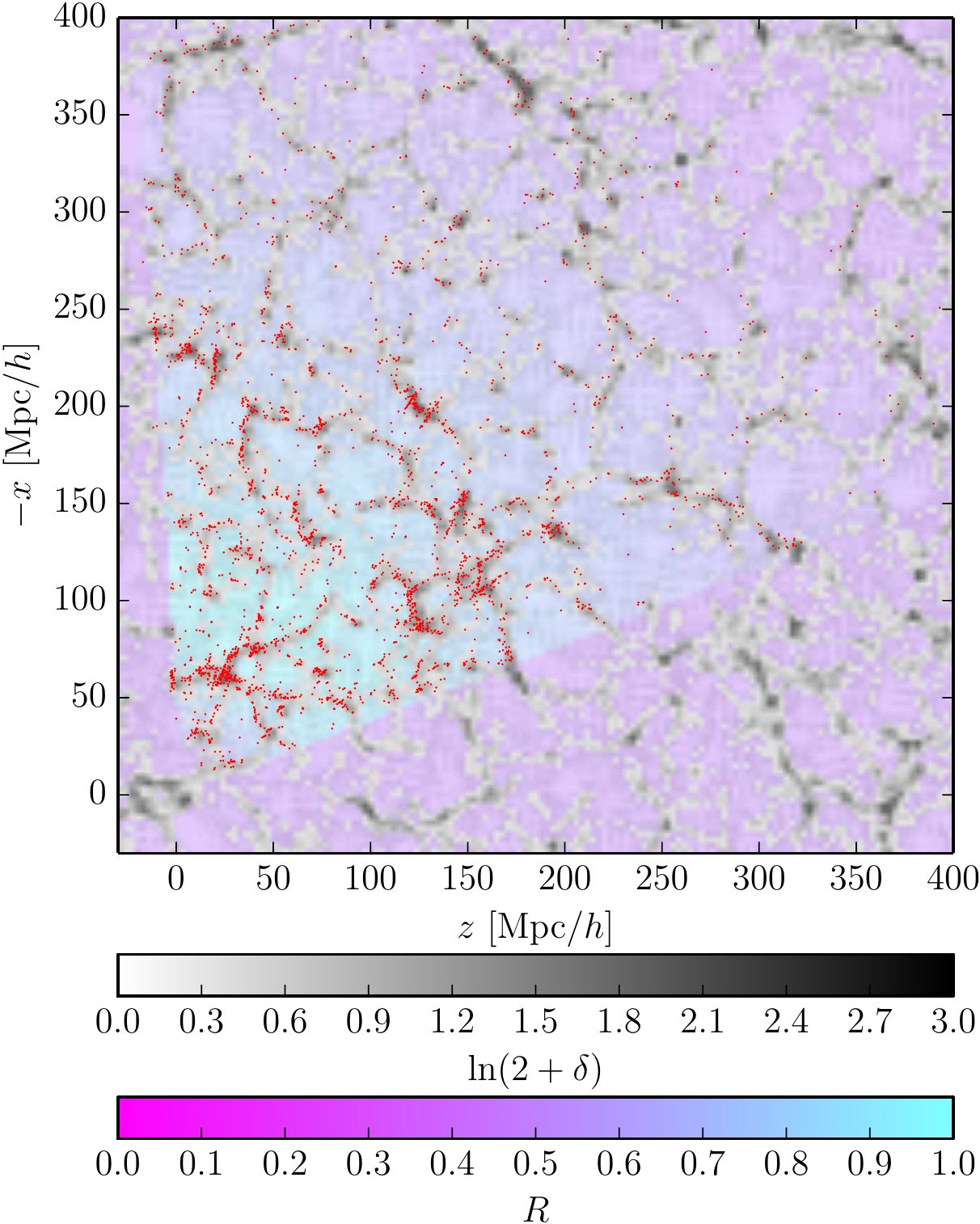}}
  {\includegraphics[type=pdf,ext=.pdf,read=.pdf,width=\columnwidth]{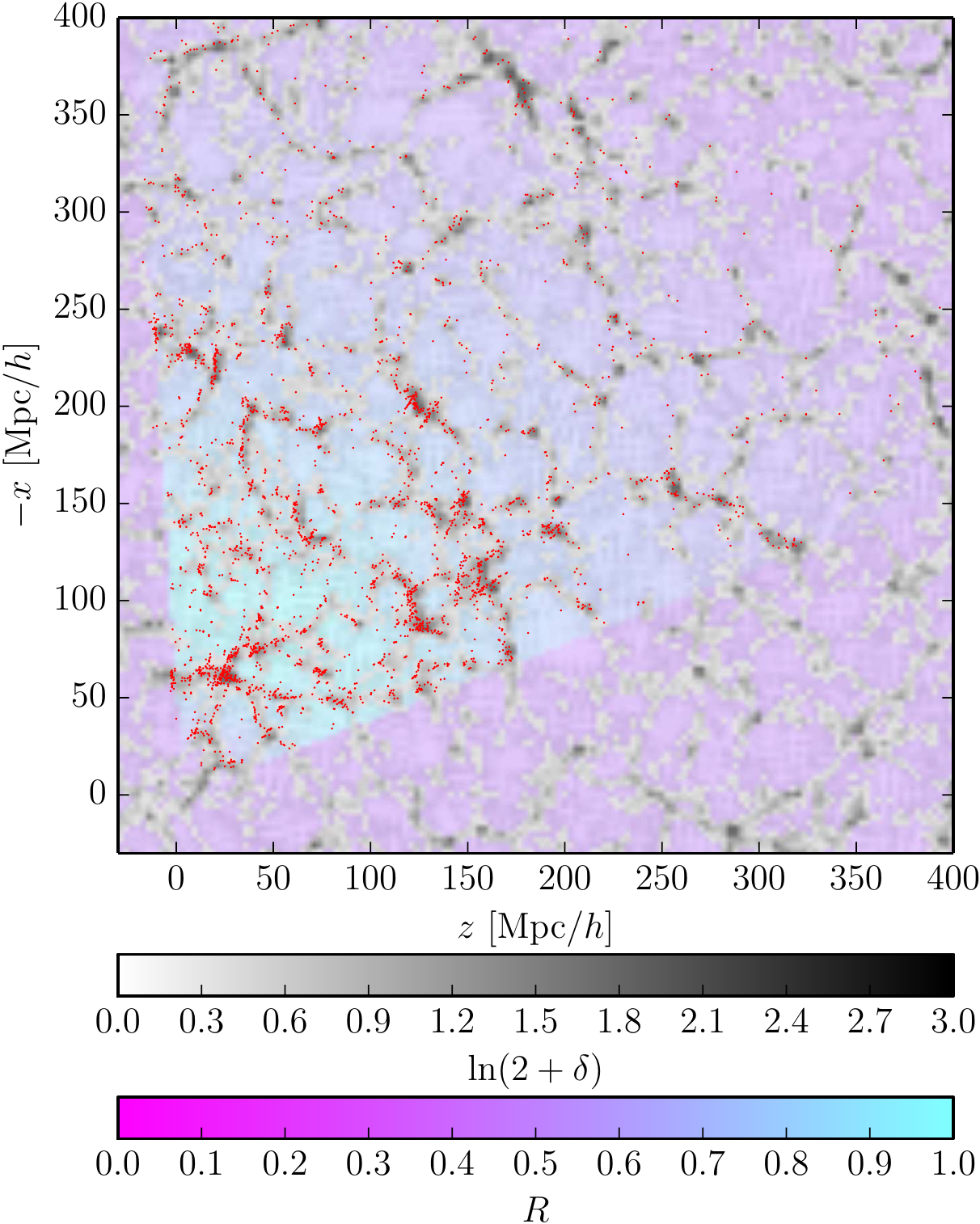}}
  \caption{
Slices through different data-constrained realizations used to build samples of the dark matter void catalog. The SDSS galaxies used for the inference with {\borg} are represented as red dots. The density of dark matter particles identified by {\vide} as being part of a void is shown in gray scale. In color, we show the particles that live in the core of dark matter voids (in a density environment smaller than $-0.3$ times the average density). The survey response operator $R$ shows how well the results are constrained by the data (see text for details). In the observed region, the data are strongly informative about the cosmic web in general and voids in particular; the reconstructions are not prior-dominated.
           }
\label{fig:slice_voids}
\end{figure*}

In figure \ref{fig:slice_voids}, we show slices through different data-constrained realizations. The density of dark matter particles identified by {\vide} as being part of a void is represented in gray scale. Note that, since $\textsc{zobov}$ essentially performs a division of space in different void regions with vanishingly-thin ridges, almost all particles initially present in the dark matter field are conserved. For clarity of the visualization, the quantity represented is $\ln(2+\delta)$ where $\delta$ is the density contrast of particles in voids. The SDSS galaxies used for the {\borg} analysis are overplotted as red dots. The core of dark matter voids (using a density threshold $\delta < -0.3$) is shown in color. As can be observed, dark matter voids also correspond to underdensities in the field traced by galaxies, which is in agreement with the results obtained by \cite{Sutter2014DMGALAXYVOIDS} in simulations.

\subsubsection{Selection of voids}
\label{sec:Selection of voids}

The {\vide} pipeline identifies all dark matter voids in the non-linear data-constrained realizations described in section \ref{sec:Generation of data-constrained reconstructions}. These live in boxes of 750 Mpc/$h$ side length with periodic boundary conditions. In order to select physically meaningful dark matter void candidates, we have to select a subsample of voids which intersect the volume of the box actually constrained by SDSS galaxies.

As described in \citetalias{Jasche2015BORGSDSS}, unobserved and observed regions in the inferred final density fields do not appear visually distinct, a consequence of the excellent performance of the 2LPT model implemented in {\borg} as a physical description of structure formation. In addition, due to the non-local transport of observational information between initial and final conditions, the region influenced by data extends beyond the survey boundaries and the large-scale structure appears continuous there. The fact that data constraints can radiate out of the survey volume has been known since the first constrained reconstructions of the mass distribution \citep{Bertschinger1987,Hoffman1991,vandeWeygaert1996}, where a power spectrum prior was assumed to sample constrained Gaussian random fields. Here, as detailed in \citetalias{Jasche2015BORGSDSS}, constraints are propagated by the structure formation model assumed in the inference process (2LPT), which accounts not only for two-point statistics, but for the full hierarchy of correlators, in its regime of validity. Therefore, dark matter voids candidates intersecting the survey boundaries can be considered as physical if a significant fraction of their volume is influenced by the data.

The survey response operator $R$ is a voxel-wise function representing simultaneously the survey geometry (observed and unobserved regions) and the selection effects in galaxy catalogs. Here, we kept for $R$ the average over the six luminosity bins used in the {\borg} SDSS run \citepalias[for details see][]{Jasche2015BORGSDSS}. For the purpose of this work, we keep all void candidates whose center is in a region where $R$ is strictly positive. This region represents $7.9 \times 10^7$ cubic Mpc/$h$, around 18.7\% of the full box. In each of the 11 realizations used in this work, we kept around 166,000 data-constrained voids out of 886,000 voids in the entire box.

In figure \ref{fig:slice_voids}, the survey response operator is shown in color from purple (totally unobserved region) to blue (region fully constrained by the data). One can see the correct propagation of information operated by {\borg}, as voids appear continuous at the survey boundaries.

\subsection{Blackwell-Rao estimators for dark matter void realizations}
\label{sec:Blackwell-Rao estimators for dark matter void realizations}

A particular advantage of our Bayesian methodology is the ability to provide accurate uncertainty quantification for derived dark matter void properties. In particular, the Markovian samples provided by \citetalias{Jasche2015BORGSDSS} permit us to employ a Blackwell-Rao estimator to describe the posterior distribution for inferred dark matter voids. Specifically, we are interested in deriving the posterior distribution  $\mathcal{P}(x \vert d)$ of a dark matter void property $x$ given observations $d$. Using the realizations of the initial conditions $\delta^\mathrm{i}$ taken from \citetalias{Jasche2015BORGSDSS} and the dark matter void realizations $V$ generated by the approach described in sections \ref{sec:Generation of data-constrained reconstructions} and \ref{sec:Void finding and processing}, we obtain

\begin{eqnarray}
\mathcal{P}(x \vert d) &=& \int \mathcal{P}(x \vert V) \, \mathcal{P}(V ,\delta^{\mathrm{i}} \vert d) \, \mathrm{d}V \, \mathrm{d} \delta^{\mathrm{i}} \nonumber \\
&=& \int \mathcal{P}(x \vert V) \, \mathcal{P}(V \vert \delta^{\mathrm{i}},d) \, \mathcal{P}(\delta^{\mathrm{i}} \vert d) \, \mathrm{d}V \, \mathrm{d} \delta^{\mathrm{i}} \nonumber \\
&=& \int \mathcal{P}(x \vert V) \, \delta^\mathrm{D} (V - \tilde{V}( \delta^{\mathrm{i}} )) \, \mathcal{P}(\delta^{\mathrm{i}} \vert d) \, \mathrm{d}V \, \mathrm{d} \delta^{\mathrm{i}} \nonumber \\
&=& \int \mathcal{P}(x \vert \tilde{V}( \delta^{\mathrm{i}} )) \,\mathcal{P}(\delta^{\mathrm{i}} \vert d)\, \mathrm{d} \delta^{\mathrm{i}} \nonumber \\
&\approx& \frac{1}{N} \sum_k \mathcal{P}\left(x \vert \tilde{V}( \delta^{\mathrm{i}}_k )\right) \nonumber \\
&=& \frac{1}{N} \sum_k \mathcal{P}\left(x \vert V_k\right) \, ,
\label{eq:bwrao}
\end{eqnarray}
where we assumed the dark matter void templates $V$ to be conditionally independent of the data $d$ given the initial conditions $\delta^\mathrm{i}$, and to derive uniquely from the initial density field via the procedure described in sections \ref{sec:Generation of data-constrained reconstructions} and \ref{sec:Void finding and processing}, yielding $\mathcal{P}(V \vert \delta^{\mathrm{i}}, d) = \mathcal{P}(V \vert \delta^{\mathrm{i}}) = \delta^\mathrm{D} (V - \tilde{V}( \delta^{\mathrm{i}}))$. We also exploited the fact that we have a sampled representation of the initial conditions posterior distribution $\mathcal{P}(\delta^{\mathrm{i}} \vert d) \approx 1/N \sum_k \delta^\mathrm{D}(\delta^{\mathrm{i}}-\delta^{\mathrm{i}}_k )$, where $k$ labels one of the $N$ samples. The last line of equation \eqref{eq:bwrao} represents the Blackwell-Rao estimator for void property $x$ to be inferred from our dark matter void catalogs $V_k$, providing thorough Bayesian means to quantify uncertainties. It consists of a mixture distribution over different realizations of dark matter void templates.

The {\vide} pipeline provides estimated means and variances for derived quantities \(x\), allowing us to model the distributions $\mathcal{P}(x \vert V_k)$ as Gaussians with mean $x_k$ and variance $\sigma_k^2$, for respective dark matter void templates. The final expression for the posterior distribution of $x$ given the data is therefore
\begin{equation}
\label{eq:bwrao_estimator}
\mathcal{P}(x \vert d) \approx \frac{1}{N} \sum_k \frac{1}{\sqrt{2\,\pi\,\sigma_k^2}} \, \exp \left(-\frac{1}{2} \frac{\left( x - x_k \right)^2}{\sigma_k^2} \right).
\end{equation}
Even though we have access to non-Gaussian uncertainty quantification via the posterior distribution given in equation \eqref{eq:bwrao_estimator}, for the presentation in this paper we will be content with estimating means and variances. 
The mean for $x$ given $d$ is 
\begin{equation}
\label{eq:BRmean}
\langle x \vert d \rangle \approx \frac{1}{N} \sum_k x_k , 
\end{equation}
and the variance is
\begin{equation}
\label{eq:BRvar}
\langle \left( x - \langle x \rangle \right)^2 \vert d \rangle \approx \frac{1}{N} \sum_k (x_k^2 + \sigma_k^2) - \langle x \vert d \rangle^2 .
\end{equation}

As described in section \ref{sec:Selection of voids}, we select voids in the data-constrained regions of reconstructions of the dark matter density field. Since these regions are the same in different reconstructions, the different void catalogs describe the same region of the actual Universe. For this reason, while estimating uncertainties, it is not possible to simply use all the voids in our catalogs as if they were independent\footnote{We generally recommend special care for proper statistical treatment while working with the data-constrained realizations of our dark matter void catalog, especially if one wants to use frequentist estimators of void properties.}. However, using an increasing number of reconstructions, we shall still see a decrease of statistical uncertainty. Indeed, from \eqref{eq:BRmean} and \eqref{eq:BRvar} it follows that
\begin{equation}
\langle \left( x - \langle x \rangle \right)^2 \vert d \rangle \leq \frac{1}{N} \sum_k \sigma_k^2 ,
\end{equation}
which means that the combination of different realizations will generally yield an improved estimator for any original statistics.

Note that this procedure is completely general and applies to any estimator provided by the {\vide} pipeline.

\subsection{Void catalogs for comparison of our results}
\label{sec:Galaxy void catalog and dark matter simulation}

In section \ref{sec:Properties of dark matter voids}, we will compare our results for dark matter voids to state-of-the-art results for galaxy voids. To do so, we use the catalogs of \cite{Sutter2012DR7CATALOGS} based on the SDSS DR7 galaxies, publicly available at \href{http://www.cosmicvoids.net}{http://www.cosmicvoids.net}. In particular, we compare to the voids found in the \texttt{bright1} and \texttt{dim1} volume-limited galaxy catalogs, for which the mean galaxy separations are 8 and 3 Mpc/$h$, respectively \cite[for details, see][]{Sutter2012DR7CATALOGS}. 

Assessment of our results for dark matter voids in SDSS data also require systematic comparison to dark matter voids found in cosmological simulations. We ran 11 such unconstrained simulations with the same setup as described in section \ref{sec:Generation of data-constrained reconstructions} for the generation of data-constrained realizations. We started from Gaussian random fields with an \cite{Eisenstein1998,Eisenstein1999} power spectrum using the fiducial cosmological parameters of the {\borg} analysis ($\Omega_\mathrm{m}~=~0.272$, $\Omega_\mathrm{\Lambda}~=~0.728$, $\Omega_\mathrm{b}~=~0.045$, $h~=~0.702$, $\sigma_8~=~0.807$, $n_\mathrm{s}~=~0.961$, see \citetalias{Jasche2015BORGSDSS}). These initial density fields, defined in a 750 Mpc/$h$ cubic box of $256^3$ voxels, are occupied by a Lagrangian lattice of $512^3$ dark matter particles. These are evolved to $z=69$ with 2LPT and from $z=69$ to $z=0$ with \textsc{Gadget-2}. As for constrained realizations, in our simulations we selected the voids located inside the observed SDSS volume (see section \ref{sec:Selection of voids}) and combined properties using Blackwell-Rao estimators (see section \ref{sec:Blackwell-Rao estimators for dark matter void realizations}).


\section{Properties of dark matter voids}
\label{sec:Properties of dark matter voids}

In this section, we describe the statistical properties of the dark matter voids found in the data-constrained parts of our reconstructions of the SDSS volume. We focus on three key statistical summaries abundantly described in the literature: number count, ellipticity distribution and radial density profiles.

\subsection{Number function}
\label{sec:Number function}

\begin{figure} 
  \centering 
  {\includegraphics[type=pdf,ext=.pdf,read=.pdf,width=\columnwidth]{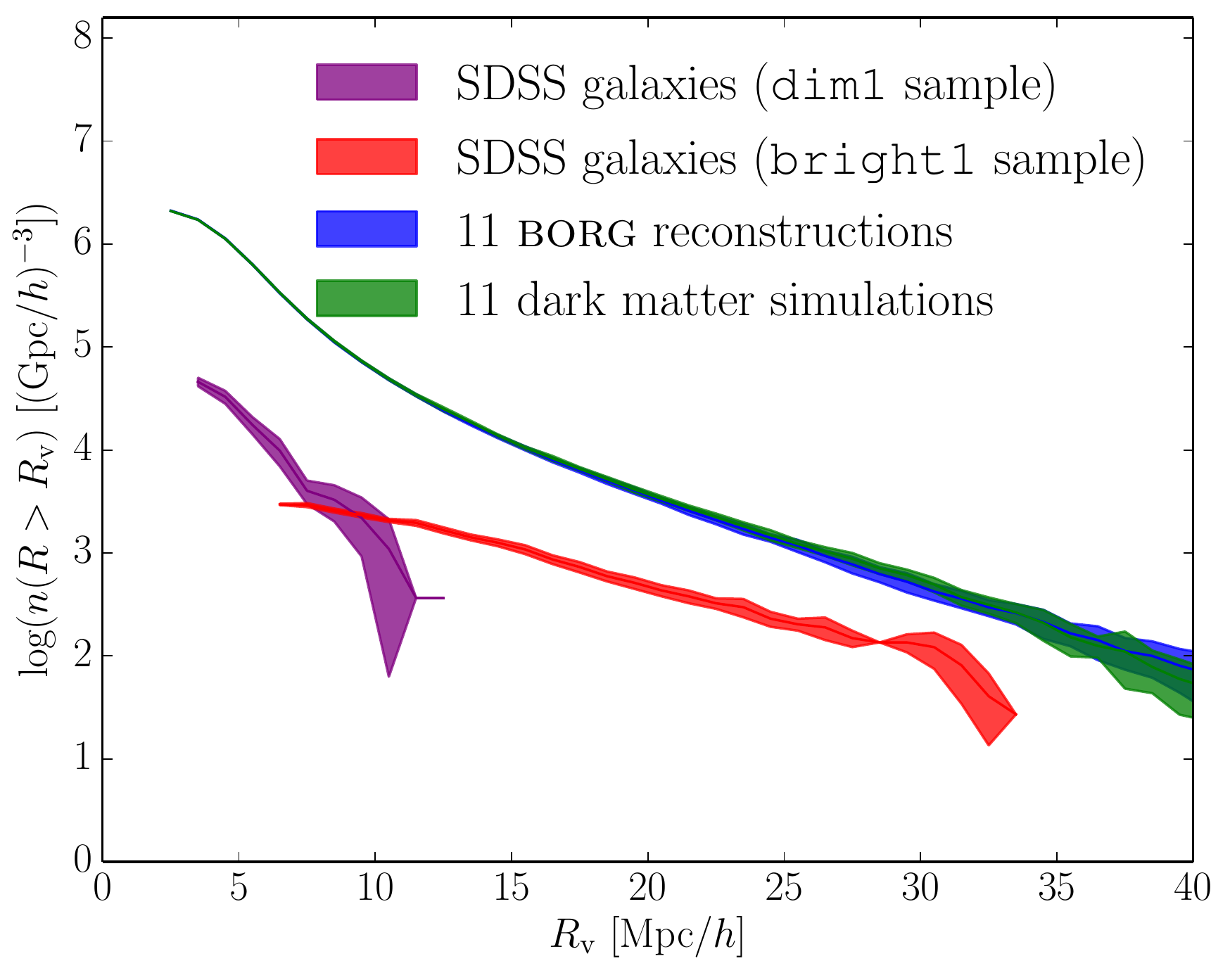}}
  \caption{
           Cumulative void number functions. The results from 11 {\borg} reconstructions (blue) are compared to a dark matter $N$-body simulation (green) and to the galaxy voids directly found in two volume-limited sub-samples of the SDSS DR7 (\texttt{dim1}, purple and \texttt{bright1}, red). The solid lines are the measured or predicted number functions and the shaded regions are the 2-$\sigma$ Poisson uncertainties. Fewer voids are found in observations than in dark matter simulations, due to the sparsity and bias of tracers, as well as observational uncertainty coming from the survey geometry and selection effects. Number functions from {\borg} reconstructions agree with simulations at all scales.}
\label{fig:numberfunc}
\end{figure}

The number function of voids provides a simple, easily accessible, and surprisingly sensitive cosmological probe. For example, the number function has been shown to respond to coupled dark matter-dark energy \citep{Li2009,Sutter2015COUPLEDDEDM}, modified gravity \citep{Li2012,Clampitt2013}, and variations in fundamental cosmological parameters \citep{Pisani2015}. While most studies of the number function take place in $N$-body simulations, there has also been significant theoretical and analytical work, beginning with the excursion set formulation of~\cite{Sheth2004} and continuing through further enhancements to account for the complex nature of void shapes~\citep{Jennings2013}. As previous authors \citep{Muller2000,Sutter2012DR7CATALOGS,Sutter2014DR9CATALOGS,Nadathur2014a,Nadathur2014b} have noted, there tend to be fewer voids in observations than in numerical simulations, especially for small voids. This is due to the conjugate effects of sparsity and biasing of tracers, which can modify the number function in complex ways \citep{Furlanetto2006,Sutter2014SPARSITYBIAS,Sutter2014DR9CATALOGS}, as well as survey geometries and selection effects, which can non-trivially diminish the void population. However, recently~\cite{Sutter2014DR9CATALOGS} showed a correspondence between observed and theoretical number functions once these factors are taken into account.

Figure \ref{fig:numberfunc} shows the cumulative void number function in {\borg} reconstructions (blue) compared to dark matter simulations using the same setup (green) and to galaxy voids in the SDSS DR7 (red and purple). The confidence regions are 2-$\sigma$ Poisson uncertainties and the blue and green lines use Blackwell-Rao estimators to combine the results in 11 realizations.

We can immediately note the excellent agreement between simulations and dark matter voids candidates in the SDSS as found by our methodology. The two void populations are almost indistinguishable at all scales, which demonstrates that the data-constrained number function predicted by our methodology is exactly that of dark matter voids in numerical simulations. In particular, this proves that our framework correctly permits to circumvent the effects of sparsity and biasing of SDSS galaxies on void number count. Indeed, dark matter voids in our reconstructions are densely-sampled with the same number density as in simulations, $\bar{n}~=~0.318~(\mathrm{Mpc}/h)^{-3}$ ($512^3$ particles in $(750~\mathrm{Mpc}/h)^3$) compared to $\bar{n}~\approx~10^{-3}~(\mathrm{Mpc}/h)^{-3}$ for SDSS galaxies \citep{Sutter2012DR7CATALOGS}. Furthermore, any incorrect treatment of galaxy bias by the {\borg} algorithm would result in a residual bias in our reconstructions that would yield an erroneous void number function as compared to simulations \citep{Sutter2014SPARSITYBIAS}. The absence of any such feature confirms that galaxy bias is correctly accounted for in our analysis and further validates the framework described in \citetalias{Jasche2015BORGSDSS}.

Additionally, due to the high density of tracer particles, we find at least around one order of magnitude more voids at all scales than the voids directly traced by the SDSS galaxies, which sample the underlying mass distribution only sparsely. This results in a drastic reduction of statistical uncertainty in void catalogs, as we demonstrate in sections \ref{sec:Ellipticity distribution} and \ref{sec:Radial density profiles}.

\subsection{Ellipticity distribution}
\label{sec:Ellipticity distribution}

The shape distribution of voids is complementary to overdense probes of the dark matter density field such as galaxy clusters. Indeed, as matter collapses to form galaxies, voids expand and can do so aspherically. While \citet{Icke1984} argued that voids are expected to become more spherical as they expand, \citet{Platen2008} found that the shape distribution of voids remains complex at late times and showed that the aspherical expansion of voids is strongly linked to the external tidal influence\footnote{Tidal effects are taken into account in our analysis since {{\borg}} models gravitational evolution up to second order in Lagrangian perturbation theory.}. Therefore, the shapes of empty regions generally change during cosmic evolution and retain information on their formation history. In particular, the void shape distribution potentially serves as a powerful tracer of the equation of state of dark energy \citep{Lee2006,Park2007,Biswas2010,Lavaux2012,Bos2012}. In addition, the mean stretch of voids along the line of sight may be used for an application of the Alcock-Paczynski test \citep{Alcock1979,Ryden1995,Lavaux2012,Sutter2012APSDSS,Sutter2014APSDSS,Hamaus2014c}.

For these applications, it is of crucial importance for the void catalog to be unaffected by systematics due to baryonic physics. Furthermore, as pointed out by \cite{Bos2012}, in sparse populations such as galaxies it is very difficult to statistically separate $\Lambda$CDM from alternative cosmologies using void shapes. As we now show, our framework allows to access void shapes at the level of the dark matter distribution, deeper than with the galaxies, and to reduce the statistical uncertainty due to their sparsity. Note that all the phase information and spatial organization of the LSS is unaffected by our prior assumptions, which generally affect the density amplitudes via the cosmological power spectrum. The geometry of voids discussed here is therefore strongly constrained by the observations.

\begin{figure} 
  \centering 
  {\includegraphics[type=pdf,ext=.pdf,read=.pdf,width=\columnwidth]{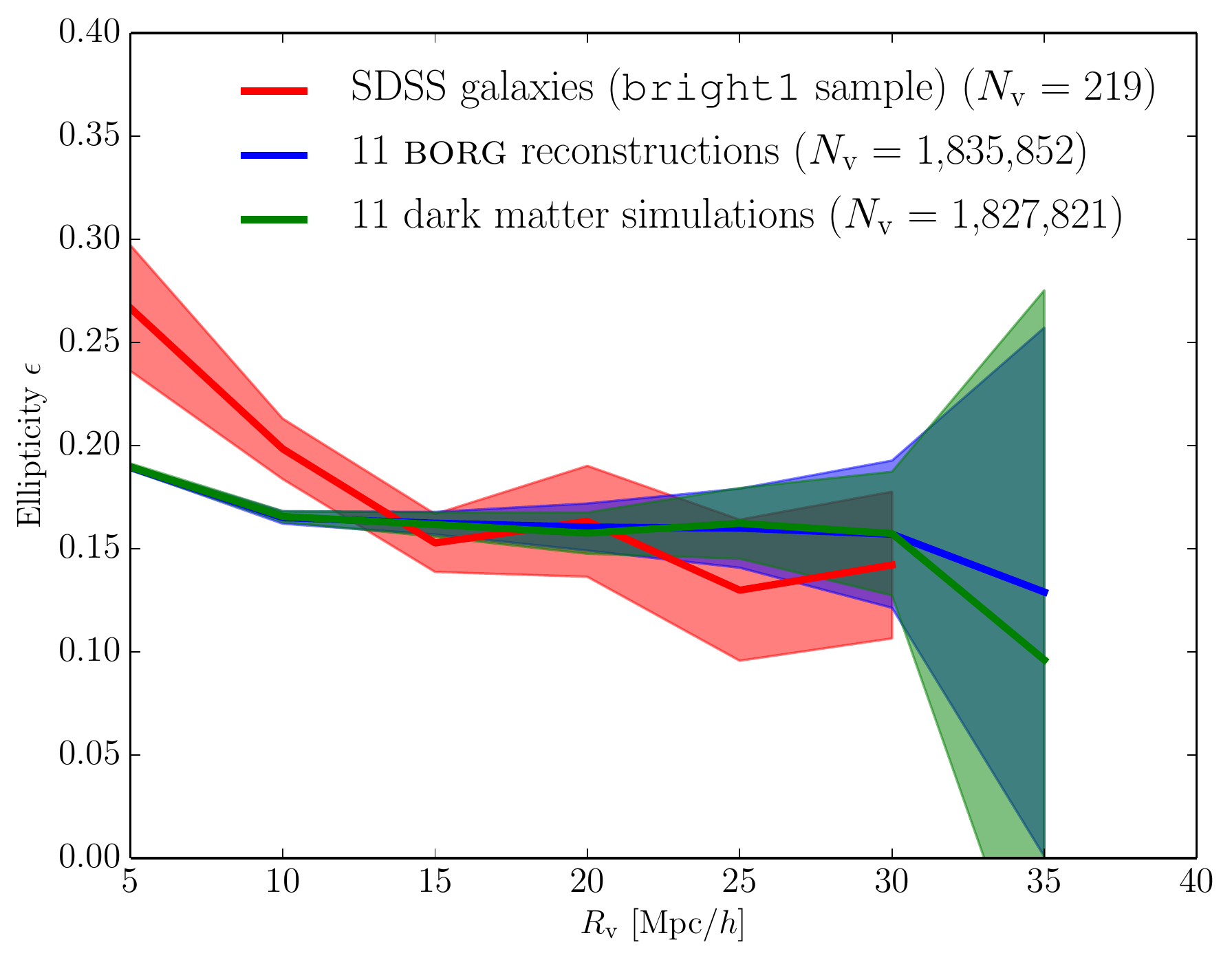}}
  \caption{
           Distribution of ellipticities $\epsilon$ versus effective radii of voids. The solid line shows the mean, and the shaded region is the 2-$\sigma$ confidence region estimated from the standard error on the mean in each radial bin. Small galaxy voids are found more elliptical than dark matter voids because of important Poisson fluctuations below the mean galaxy separation ($8~\mathrm{Mpc}/h$). Ellipticities of dark matter voids in {\borg} reconstructions and simulations agree at all scales, and the statistical uncertainty in their determination is drastically reduced in comparison to galaxy void catalogs.
           }
\label{fig:ellipvsr}
\end{figure}

We simplify the discussion by focusing on the ellipticity, computed by the {\vide} toolkit using the eigenvalues of the inertia tensor \citepalias[for details, see][]{Sutter2015VIDE}. Figure \ref{fig:ellipvsr} shows the mean ellipticity and the standard error on the mean (i.e. $\sigma/\sqrt{N_\mathrm{v}}$, where $\sigma$ is the standard deviation and $N_\mathrm{v}$ is the number of voids) as a function of void effective radius. The red line represents the galaxy voids directly found in the SDSS data, the blue line the dark matter voids of our data-constrained catalogs, and the green line the voids found in dark matter simulations prepared with the same setup. The blue and green lines use Blackwell-Rao estimators to combine the results of 11 realizations. For the interpretation of the ellipticity of small galaxy voids, it is useful to recall that the mean galaxy separation in the \texttt{bright1} sample is 8~Mpc/$h$, meaning that Poisson fluctuations will be of importance for voids whose effective radius is below this scale. 

The comparison between dark matter voids of {\borg} reconstructions and of simulations shows that the predicted ellipticities fully agree with the expectations at all scales. This further demonstrates that our candidates qualify as dark matter voids as defined by numerical simulations, in particular alleviating the galaxy bias problem. Furthermore, as already noted, our inference framework produces many more voids than sparse galaxy catalogs, especially at small scales. This results in a radical reduction of statistical uncertainty in the ellipticity prediction for small dark matter voids as compared to galaxy voids, as can be observed in figure \ref{fig:ellipvsr}.

\subsection{Radial density profiles}
\label{sec:Radial density profiles}

\begin{figure*} 
  \centering 
  {\includegraphics[type=pdf,ext=.pdf,read=.pdf,width=\columnwidth]{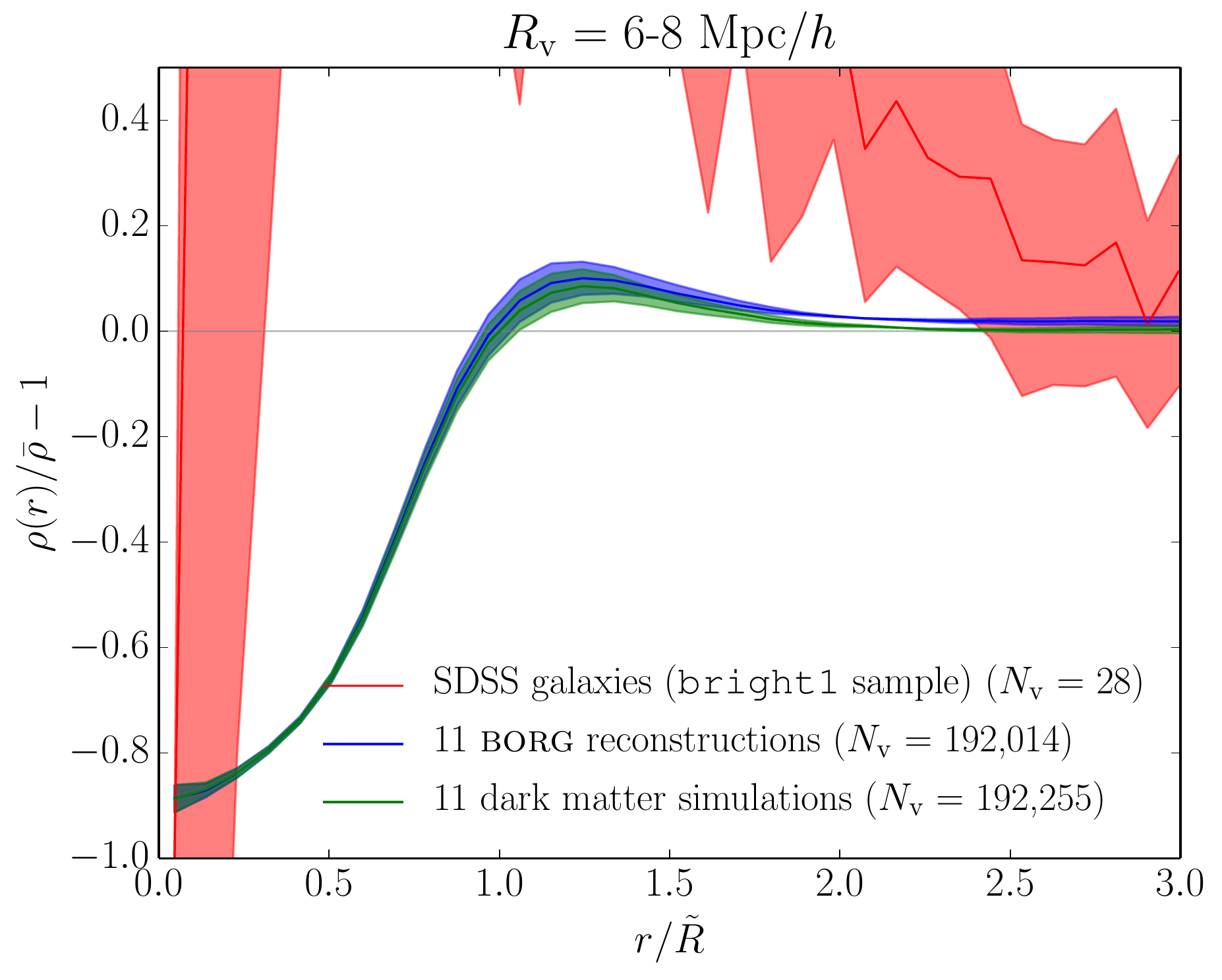}}
  {\includegraphics[type=pdf,ext=.pdf,read=.pdf,width=\columnwidth]{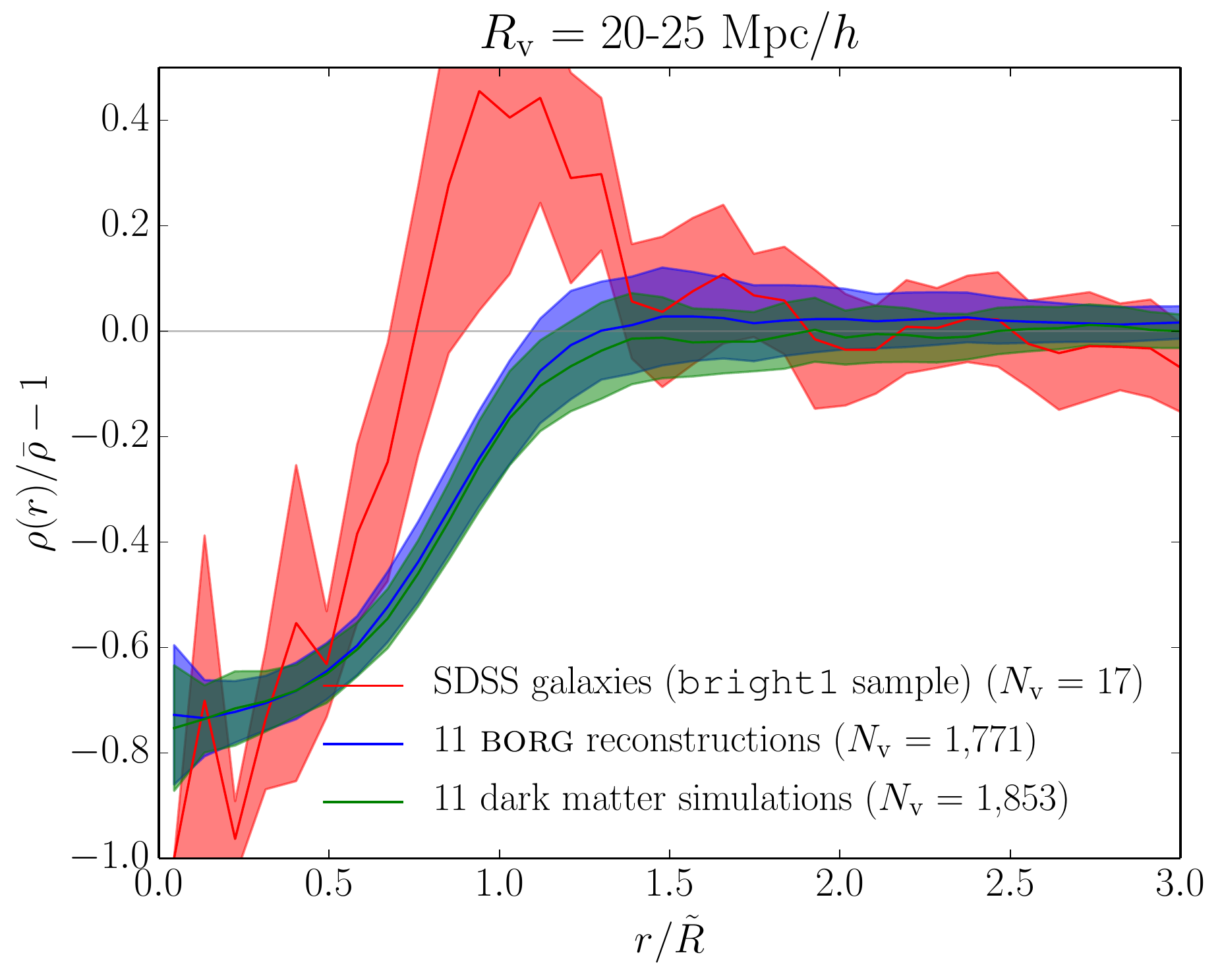}}
  \caption{
           One-dimensional radial density profiles of stacked voids, for voids of effective radius in the range 6-8~Mpc/$h$ (left) and 20-25~Mpc/$h$ (right). $\tilde{R}$ corresponds to the median void size in the stack. The solid line shows the mean, and the shaded region is the 2-$\sigma$ confidence region estimated from the standard error on the mean in each radial bin. Galaxy void profiles are strongly noise-dominated, contrary to dark matter voids. The heights of compensation ridges are different because dark matter voids are identified in a higher density of tracers, which induces a deeper void hierarchy.
           }
\label{fig:1dprofile}
\end{figure*}

The radial density profile of voids, reconstructed in real space using techniques such as those described in \cite{Pisani2014}, can be used to test general relativity and constrain dynamical dark energy models \citep{Shoji2012,Spolyar2013}. More generally, it shows a self-similar structure \citep{Colberg2005,Ricciardelli2014,Hamaus2014b,Nadathur2014b}, and characterizes the LSS in a fundamental way \citep{vandeWeygaert1993}. All results presented in this section assume that dark matter particles in {\borg} reconstructions and in simulations live in physical space. The {\borg} algorithm automatically mitigates redshift space distortions by treating anisotropic features in the data as noise \citepalias{Jasche2015BORGSDSS}. Furthermore, as pointed out by \cite{Padilla2005}, redshift space distortions have very mild effects on void density profiles. We therefore expect our results to be robust under the transformation from real to redshift space.

Using {\vide}, we construct the one-dimensional radial density profiles of stacked voids for various void sizes. Note that we do not apply any rescaling to the void sizes as we stack. Figure \ref{fig:1dprofile} shows two such profiles, for voids of effective radius in the range 6-8~Mpc/$h$ (left panel) and 20-25~Mpc/$h$ (right panel). The solid lines show the mean and the shaded regions are the 2-$\sigma$ confidence regions estimated from the standard error on the mean, using Blackwell-Rao estimators for {\borg} reconstructions and dark matter simulations. At the level of statistical error in our results, our reconstructions show radial density profiles in agreement with simulations at all radii and for all void sizes. Note that, if small voids essentially reflect the prior information used for the {\borg} analysis and $N$-body filtering, bigger voids are strongly constrained by the data. The profile shapes agree nicely with the results of \cite{Sutter2014SPARSITYBIAS,Hamaus2014b} from dark matter simulations: higher ridges and lower central densities in smaller voids. Specifically, our reconstructions exhibit the same behaviour as simulations, with a transition scale between small overcompensated to large undercompensated voids \citep{Ceccarelli2013,Paz2013,Cai2014,Hamaus2014a}.

In contrast, galaxy void profiles at the same scales are strongly noise-dominated. This is due to the sparsity and biasing of galaxies, which are alleviated with the present approach. In particular, our methodology performs a meaningful compromise between data and prior information, which predicts corrected shapes and smaller variance for the profiles of dark matter voids as compared to galaxy voids. Note that at the same physical scales (e.g. $20~\mathrm{Mpc}/h$), galaxy voids and dark matter voids have different ridge heights. This is because a deeper void hierarchy emerges in higher tracer sampling densities, affecting the compensation of voids at a given size \citep{Sutter2014SPARSITYBIAS}. 

In addition to the location of all dark matter particles, our inference framework also provides their individual velocity vectors, which are predicted from gravitational clustering. While the direct measurement of individual galaxy velocities is very difficult in most observations, our reconstruction technique readily allows to infer the velocity profile of voids. This allows to make a connection between a static (based on the density profiles) and a dynamic (based on the velocity profiles) characterization of voids. In particular, as mentioned before, our results agree with the existence of a transition scale between two regimes: undercompensated, inflowing voids and overcompensated, outflowing voids, respectively known as void-in-cloud and void-in-void in the terminology originally introduced by \cite{Sheth2004}.


\section{Summary and conclusions}
\label{sec:Summary and conclusions}

This paper is an example of the rich variety of scientific results that have been produced by the recent application \citepalias{Jasche2015BORGSDSS} of the Bayesian inference framework {\borg} \citepalias{Jasche2013BORG} to the Sloan Digital Sky Survey main sample galaxies. We proposed a method designed to find dark matter void candidates in the Sloan volume. In doing so, we proved that physical voids in the dark matter distribution can be correctly identified by the \textit{ab initio} analysis of galaxy surveys. 

Our method relies on the physical inference of the initial conditions for the entire LSS \citepalias{Jasche2013BORG,Jasche2015BORGSDSS}. Starting from these, we generated realizations of the LSS using a fully non-linear cosmological code. In this fashion, as described in section \ref{sec:Generation of data-constrained reconstructions}, we obtained a set of data-constrained reconstructions of the present-day dark matter distribution. The use of fully non-linear dynamics as a filter allowed us to extrapolate the predictions of {\borg} to the unconstrained non-linear regimes and to obtain an accurate description of structures. As described in section \ref{sec:Void finding and processing}, we identified the voids in these reconstructions using the void finder of the {\vide} pipeline \citepalias{Sutter2015VIDE} and applied an additional selection criterion to limit the final catalogs of dark matter voids candidates to regions covered by observations. To check that these candidates qualify for physical voids, we analyzed our catalogs in terms of a set of statistical diagnostics. We focused on three key void statistics, well understood both in data and in simulations, provided by the {\vide} toolkit: number function, ellipticity distribution and radial density profile. As mentioned in section \ref{sec:Galaxy void catalog and dark matter simulation}, for comparison of our results, we used the void catalog of \cite{Sutter2012DR7CATALOGS}, directly based on SDSS main sample galaxies, and unconstrained dark matter simulations produced with the same setup as our reconstructions.

For quantifying the uncertainty, we adopted the same Bayesian philosophy as in the LSS inference framework: several void catalogs are produced, based on different samples of the {\borg} posterior probability distribution function. Each of them represents a realization of the actual dark matter voids in the Sloan volume, and the variation between these catalogs quantifies the remaining uncertainties of various sources (in particular, survey geometry and selection effects, see \citetalias{Jasche2015BORGSDSS} for a complete discussion). In order to produce a statistically meaningful combination of our different dark matter void catalogs, in section \ref{sec:Blackwell-Rao estimators for dark matter void realizations}, we introduced Blackwell-Rao estimators. We showed that the combination of different realizations generally yields an improved estimator for any original void statistic.

For all usual void statistics (number function in section \ref{sec:Number function}, ellipticity distribution in section \ref{sec:Ellipticity distribution} and radial density profiles in section \ref{sec:Radial density profiles}), we found remarkably good agreement between predictions for dark matter voids in our reconstructions and expectations from numerical simulations. This validates our inference framework and qualifies the candidates to physically reasonable dark matter voids, probing a level deeper in the mass distribution hierarchy than galaxies. Further, since sparsity and biasing of tracers modify these statistics \citep{Sutter2014SPARSITYBIAS}, it means that these effects have been correctly accounted for in our analysis. Indeed, in \citetalias{Jasche2015BORGSDSS} we showed that {\borg} accurately accounts for luminosity-dependent galaxy bias and performs automatic calibration of the noise level within a fully Bayesian approach. Building on the detailed representation of initial density fields, our reconstructions possess a high density of tracers, $\bar{n} = 0.318~(\mathrm{Mpc}/h)^{-3}$, contrary to galaxies, which sample the underlying mass distribution only sparsely ($\bar{n} \approx 10^{-3}~(\mathrm{Mpc}/h)^{-3}$). 

Another important aspect of our methodology is that the use of full-scale physical density fields instead of a scarce population of galaxies allows to adjust the density of tracers to reduce shot-noise at the desired level. In our analysis, we found at least one order of magnitude more voids at all scales. This yields a radical reduction of statistical uncertainty in noise-dominated void catalogs, as we have shown for ellipticity distributions and density profiles.

In summary, our methodology permits to alleviate the issues due to the conjugate and intricate effects of sparsity and biasing on galaxy void catalogs, to drastically reduce statistical uncertainty in void statistics, and yields new catalogs of dark matter voids for a variety of cosmological applications. For example, these enhanced data sets can be used for cross-correlation with other cosmological probes such as the cosmic microwave background, to study the integrated Sachs-Wolfe effect, or gravitational lensing shear maps. Along with the ensemble mean density field and corresponding standard deviations inferred by {\borg}, published as supplementary material with \citepalias{Jasche2015BORGSDSS}, we believe that the catalogs of our dark matter voids candidates in the Sloan volume can be of interest to the scientific community. For this reason, all the void catalogs used to produce the results described in this paper will be made publicly available at \href{http://www.cosmicvoids.net}{http://www.cosmicvoids.net}.

Our Bayesian methodology, based on inference with {\borg} and subsequent non-linear filtering of the results, assumes some prior information, namely the standard $\Lambda$CDM cosmological framework and initially Gaussian density fluctuations. We want to emphasize that any analysis using our constrained catalogs will be biased toward the confirmation of these assumptions. Therefore, this method will be only applicable if the data contain sufficient support for the presence of non-standard cosmology to overrule the preference for $\Lambda$CDM and Gaussianity in our prior. However, any significant departure from standard cosmology means that the prior has been overridden by the likelihood and that such deviations really are supported by the data.

While the recommendations of \cite{Sutter2014SPARSITYBIAS} for quantifying and disentangling the effects of sparsity and biasing depend on specific survey details, our inference framework is extremely general. It allows to translate void statistics from current and future galaxy surveys to theory-like, high-resolution dark matter predictions. In this fashion, it is straightforward to decide if any particular void statistic can be directly informative about cosmology. These results indicate a new promising path towards effective and precise void cosmology at the level of the dark matter field.


\acknowledgments

We thank Guilhem Lavaux and Alice Pisani for valuable discussions and help with the {\vide} pipeline, and all participants in the ``Cosmic Voids in the Next Generation of Galaxy Surveys'' Workshop held in Columbus, Ohio, for insightful comments. We are grateful to the referee for a very constructive report which helped finalizing this work. Special thanks go to St\'ephane Rouberol for his support, in particular for guaranteeing flawless use of all required computational resources. The results described in this paper have been presented at the 49th ``Rencontres de Moriond'' Cosmology Session and at the International Astronomical Union Symposium 306 ``Statistical Challenges in 21st Century Cosmology''. We would like to express our gratitude to the organizers. FL acknowledges funding from an AMX grant (\'Ecole polytechnique ParisTech). JJ is partially supported by a Feodor Lynen Fellowship by the Alexander von Humboldt foundation and BW's Chaire d'Excellence from the Agence Nationale de la Recherche (ANR-10-CEXC-004-01). PMS is supported by the INFN IS PD51 ``Indark''. This work has been done within the Labex ILP (reference ANR-10-LABX-63) part of the Idex SUPER, and received financial state aid managed by the Agence Nationale de la Recherche, as part of the programme Investissements d'avenir under the reference ANR-11-IDEX-0004-02.

\bibliography{DMVoidsInSDSS}

\end{document}